\documentclass[12pt]{iopart}
\usepackage{inputenc}
\usepackage{graphicx}
\usepackage{subfigure}
\usepackage{color}
\usepackage{cite}
\usepackage{setspace}
\usepackage{cases}
\usepackage{braket}
\usepackage{multirow}

\begin{document}

\title[A Simple, Versatile Laser System for the Creation of Ground State Molecules]{A Simple, Versatile Laser System for the Creation of Ultracold Ground State Molecules}

\author{P~D~Gregory, P~K~Molony, M~P~K\"{o}ppinger, A~Kumar, Z~Ji, B~Lu, A~L~Marchant, S~L~Cornish}

\address{Joint Quantum Centre (JQC) Durham-Newcastle, Department of
Physics, Durham University, South Road, Durham DH1 3LE, United Kingdom}


\ead{s.l.cornish@durham.ac.uk}

\begin{abstract}

A narrow-linewidth, dual-wavelength laser system is vital for the creation of ultracold ground state molecules via stimulated Raman adiabatic passage (STIRAP) from a weakly bound Feshbach state. Here we describe how a relatively simple apparatus consisting of a single fixed-length optical cavity can be used to narrow the linewidth of the two different wavelength lasers required for STIRAP simultaneously. The frequency of each of these lasers is referenced to the cavity and is continuously tunable away from the cavity modes through the use of non-resonant electro-optic modulators. Self-heterodyne measurements suggest the laser linewidths are reduced to several hundred Hz. In the context of $^{87}$Rb$^{133}$Cs molecules produced via magnetoassociation on a Feshbach resonance, we demonstrate the performance of the laser system through one- and two-photon molecular spectroscopy. Finally, we demonstrate transfer of the molecules to the rovibrational ground state using STIRAP.

\end{abstract}

\maketitle

\section{Introduction}\label{sec:Introduction}

The long-range, anisotropic dipole-dipole interactions between ultracold polar molecules offer a striking contrast to the isotropic short-range interactions usually encountered in ultracold atomic gas experiments. Such dipole-dipole interactions can be significant over a range greater than the inter-site separation in a typical optical lattice leading to a range of novel quantum phases~\cite{Wall:2009, Capogrosso-Sansone:2010, Micheli:2007}. Additionally, the option of tuning these interactions by varying an applied electric field, in combination with the exquisite control of ultracold systems promises numerous possibilities in the fields of quantum controlled chemistry~\cite{Ospelkaus:2010, Krems:2008}, precision measurement~\cite{Flambaum:2007, Isaev:2010, Hudson:2011}, quantum computation~\cite{DeMille:2002} and quantum simulation~\cite{Santos:2000, Baranov:2012}.

Due to the additional rotational and vibrational degrees of freedom, even simple diatomic molecules possess a highly complex internal structure. While this makes them interesting to study, it poses major challenges in the application of standard laser cooling techniques, though recent experimental results in this direction are encouraging~\cite{Shuman:2010, Hummon:2013, Zhelyazkova:2014}. An alternative approach is to exploit the relative simplicity and wealth of experience that surrounds the laser cooling of atoms, and apply a two-step process to associate molecules from a pre-cooled atomic sample~\cite{Kohler:2006, Jones:2006}. This can be achieved by first creating weakly-bound molecules by magnetoassociation on a Feshbach resonance~\cite{Kohler:2006, Chin:2010}, followed by the subsequent optical transfer into their rovibrational ground state via stimulated Raman adiabatic passage (STIRAP)~\cite{Bergmann:1998}. A number of groups are currently pursuing this method of molecule production~\cite{Quemener:2012} yet despite the successful application of this technique in several systems~\cite{Ni:2008, Lang:2008, Danzl:2010}, the study of dipole-dipole interactions has so far been restricted to the fermionic $^{40}$K$^{87}$Rb molecule~\cite{Ni:2008}. This system has the drawback that the exchange reaction $2\mathrm{KRb}\rightarrow \mathrm{K}_2+\mathrm{Rb}_2$ is exothermic~\cite{Zuchowski:2010}. This renders KRb molecules unstable~\cite{Ospelkaus:2010} which leads to significant molecule losses~\cite{Ni:2010}. Confinement of the molecules in a three-dimensional optical lattice eliminates this reaction however~\cite{Miranda:2011}, and has led to a series of ground-breaking insights into dipolar spin exchange reactions~\cite{Yan:2013}.

Ground-state RbCs molecules hold great promise, as they offer both collisional stability, due to endothermic exchange and trimer formation reactions~\cite{Zuchowski:2010}, and a large electric dipole moment of 1.225~D~\cite{Aymar:2005, Molony:2014, Takekoshi:2014}. In recent years, significant progress has been made in magnetoassociation from a mixture of Rb and Cs, with the creation of weakly-bound $^{87}$Rb$^{133}$Cs molecules in both Innsbruck~\cite{Pilch:2009, Takekoshi:2012} and Durham~\cite{Cho:2013, Koppinger:2014}. The Innsbruck group subsequently performed detailed one- and two-photon molecular spectroscopy of the molecules near dissociation to identify a suitable path to the rovibrational ground state~\cite{Debatin:2011}. Both groups have now reported the transfer of $^{87}$Rb$^{133}$Cs molecules to the rovibrational ground state by STIRAP~\cite{Takekoshi:2014, Molony:2014}. 

In this paper, we present and characterize a simple laser system with which we have performed STIRAP transfer to the rovibrational ground state of $^{87}$Rb$^{133}$Cs from a state near dissociation following magnetoassociation from a cold atomic mixture. We give a detailed description of the optical cavity setup used to stabilize and tune the frequencies of the two lasers required for STIRAP. We characterize the laser stability using self-heterodyne measurements which yield an estimate of the laser linewidth of 0.21(1)~kHz. In addition; referencing the laser to an optical frequency comb revealed an rms deviation in the beat frequency of 116~kHz over a 24~hour period. We then demonstrate use of the laser system by carrying out molecular spectroscopy over a wide frequency range. The identification of a suitable route for STIRAP then allows the use of this system to populate the rovibrational ground state. Finally, we discuss the current limitations to the STIRAP transfer efficiency.
\section{Experimental Requirements for Efficient STIRAP Transfer}

\begin{figure}
\centering
\includegraphics[width=0.9\textwidth]{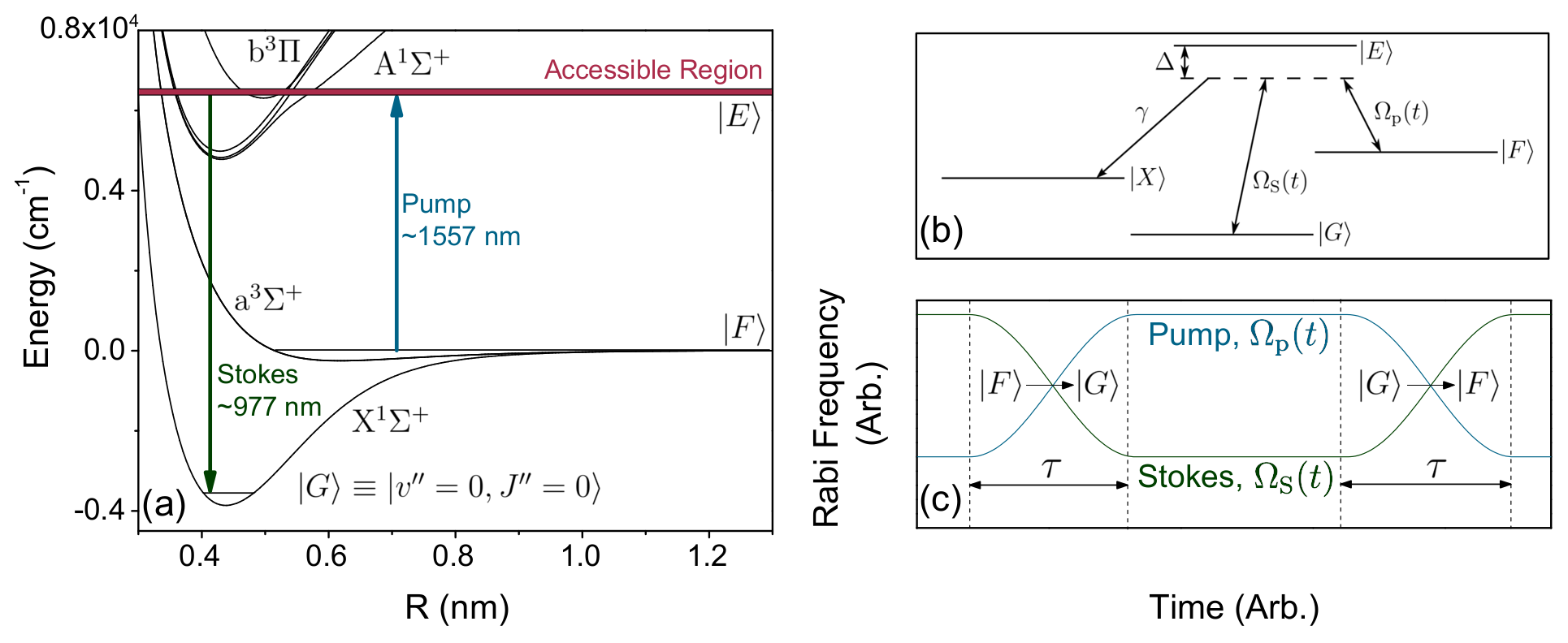}
\caption{Energy level scheme for STIRAP transfer. (a)~Potential energy curves for $^{87}$Rb$^{133}$Cs. The coupling between states introduced by the STIRAP lasers is shown. The red shaded area indicates the region in which excited states may lie and still be accessible to our laser system for coupling to both states near dissociation and the rovibrational ground state. (b)~Four energy level scheme which is used for modelling of the transfer from the initial Feshbach state~$\ket{F}$ to the ground state~$\ket{G}$ via the intermediate excited state~$\ket{E}$. Loss from the excited state is included by the introduction of a fourth dump level~$\ket{X}$ to which the excited state decays at a rate determined by the natural linewidth $\gamma$. (c)~Illustration of the STIRAP pulse sequence, used to transfer the population from state~$\ket{F}$ to state~$\ket{G}$ (and back for detection). }
\label{fig:MolecularPotentials}
\end{figure}

The large inter-atomic separation in the near-dissociation states leads to a negligible electric dipole moment and a relatively short lifetime due to collisions in the ultracold gas. This necessitates transfer of the molecules to the ground state. This is achieved by coupling both the initial weakly-bound Feshbach state~$\ket{F}$ and the rovibrational ground state~$\ket{G}\equiv\ket{v''=0, J''=0}$ to a common excited state~$\ket{E}$. This requires two lasers, hereafter referred to as the pump and Stokes lasers as shown in figure~\ref{fig:MolecularPotentials}. A suitable pulse sequence for STIRAP begins with only the Stokes light illuminating the molecules. This first counter-intuitive step initializes the molecules in a dark state $\ket{D}$ as defined by,
\begin{equation}\label{eqn:DarkState}
\ket{D} = \cos{\theta}\ket{F} - \sin{\theta}\ket{G}, ~~~~~\tan{\theta} = \frac{\Omega_{\mathrm{p}}(t)}{\Omega_{\mathrm{S}}(t)},
\end{equation} 
where $\Omega_{\mathrm{p}}(t)$ and $\Omega_{\mathrm{S}}(t)$ are the Rabi frequencies of the pump and Stokes transitions respectively. Ramping the intensity of the Stokes laser down and the pump laser up, changes these Rabi frequencies and hence the mixing angle $\theta$ which determines the composition of the dark state. In particular, with the appropriate pulse sequence the dark state can be adiabatically transformed from state~$\ket{F}$ to state~$\ket{G}$, producing molecules in the rovibrational ground state~\cite{Bergmann:1998}. Typically the sequence is then reversed (see figure~\ref{fig:MolecularPotentials}~(c)) to transfer the molecules back to the Feshbach state for dissociation and detection. 

The efficiency of STIRAP is~$100\%$ if the whole population is held in the dark state throughout the transfer. In practice however, the efficiency of the transfer ($P$) when on two-photon resonance is reduced due to non-adiabaticity of the dark state evolution, and limitations imposed by laser decoherence, such that~\cite{Yatsenko:2002}
\begin{equation}
P = \exp{\left(-\frac{\pi^{2}\gamma}{\Omega_{0}^{2}\tau} - \frac{D \tau}{2}\right)}.
\label{eqn:STIRAPEfficiency}
\end{equation}   
Here, $\gamma$ is the natural linewidth of the state~$\ket{E}$, $D$ is the linewidth associated with the frequency difference between the two lasers, $\tau$ is the transfer time (as shown in figure~\ref{fig:MolecularPotentials}~(c)) and $\Omega_{0}$ is the reduced Rabi frequency. The reduced Rabi frequency is defined as $\Omega_{0}=\sqrt{\Omega_{\mathrm{p}}^{2} + \Omega_{\mathrm{S}}^{2}}$, where $\Omega_{\mathrm{p}}$ and $\Omega_{\mathrm{S}}$ are the peak Rabi frequencies of the pump and Stokes transitions respectively. By minimizing the two contributions to the exponential in equation~\ref{eqn:STIRAPEfficiency} we are able to derive the necessary condition for efficient transfer~\cite{Aikawa:2009}:
\begin{equation} 
\frac{\Omega_{0}^{2}}{\pi^{2}\gamma} \gg \frac{1}{\tau} \gg D.
\end{equation}
The natural linewidth is dependent on the excited state chosen; the range of values for this term are therefore limited by the range of states accessible to the laser system. The importance of this term in defining the efficiency of the transfer highlights the need for a thorough molecular spectroscopy search in order to identify the best state to use, namely a state with high~$\Omega_{0}^{2}/\gamma$. This gives the first criteria which our laser system must fulfil; it must be possible to tune both of the lasers over a wide overlapping frequency range in order to maximize the range in which a suitable excited state can be found.

In practice, $\Omega_{0}^{2}/\gamma$ is limited by the available laser intensity. This sets the minimum duration for the transfer required to remain adiabatic. This in turn sets the maximum linewidth allowed to maintain coherence of the dark state. Therefore, the second criteria is that the linewidth of each of the lasers must be suitably narrow such that the linewidth associated with the frequency difference between the two lasers is minimized. In our experiment we find transitions which allow pulse durations on the order of $\sim10$~$\mu$s. This indicates that the maximum linewidth for efficient transfer must be on the order of kHz.

\section{Design of the Laser System}\label{sec:LaserSystem}

\begin{figure}
\centering
\includegraphics[width=0.9\textwidth]{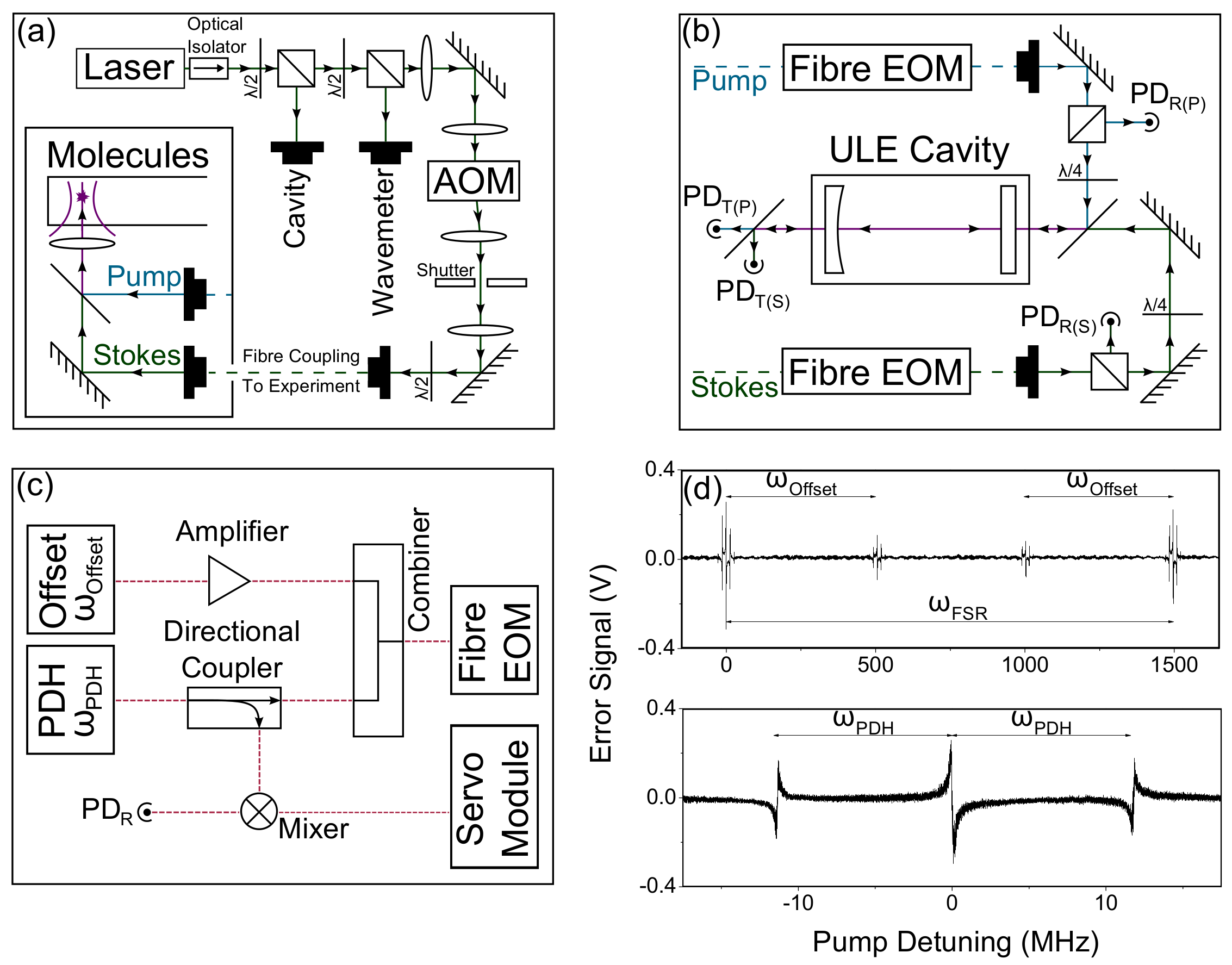}
\caption{STIRAP laser system (a)~Experimental coupling setup for the 977~nm Stokes laser. An identical setup for the 1557~nm pump laser is not shown. (b)~Optical setup for frequency stabilization to the cavity, including the fibre-coupled electro-optic modulators (EOMs) providing the Pound-Drever-Hall (PDH) and offset modulation signals.  (c)~The PDH and offset electronics for the pump laser. Note the fibre EOM provides both the PDH and offset modulations, removing the need for free space EOMs. The directional coupler which is used to split the PDH modulation signal is a Minicircuits ZDC-20-3, and the mixer is a Minicircuits ZFM-150+. The resultant error signal is sent to a Toptica FALC~110 fast analogue servo module. The Stokes setup is identical except the amplifier~(Minicircuits ZKL-1R5) is removed. (d)~PDH error signal scanning the pump laser frequency over long (upper) and short (upper) ranges.}
\label{fig:StirapLayout}
\end{figure}

A laser system for STIRAP must consist of two narrow-linewidth laser light sources. The frequency between these two sources is required to be relatively large ($\sim100$~THz) and equal to the binding energy of the molecule. This can be achieved by stabilizing the laser frequencies to an optical frequency comb~\cite{Ni:2008}, multiple independent cavities~\cite{Danzl:2010} or a single cavity~\cite{Aikawa:2011}. In the case of frequency stabilization to an optical cavity, there are two approaches. The length of the cavity may be actively stabilized by referencing back to a frequency comb~\cite{Danzl:2010}, or an atomic reference~\cite{Schunemann:1999, Debatin:2011}. Alternatively, the necessity of having an optical reference can be removed by relying on the passive stability of an ultra-low-expansion (ULE) glass cavity maintained at the zero expansion temperature of the glass~\cite{Aikawa:2011}. Typically, a tunable frequency source is then generated by using the output of another laser which is offset-locked to the frequency stabilized laser via an optical phase-locked loop~\cite{Aikawa:2011}. 

Our system utilizes a pair of Toptica DL Pro external cavity diode lasers to provide light at 1557~nm and 977~nm for the pump and Stokes transitions, respectively. Light from each laser is passed through an optical isolator ($\sim40$~dB) before being split on polarizing beam splitters and coupled into three separate fibres leading to the main experiment, a wavemeter and an optical cavity, as shown in figure~\ref{fig:StirapLayout}~(a). Both lasers are referenced to the same cylindrical 10~cm plane-concave optical cavity to narrow the linewidth. The cavity (ATFilms) is constructed from ULE glass, and is mounted in a temperature stabilized vacuum housing from Stable Laser Systems. The temperature of the cavity is maintained at 35~$^{\circ}$C, the zero-expansion temperature of the ULE glass. Further key properties of the reference cavity are listed in table~\ref{table:Cavity}. 

\begin{table}
\begin{tabular}{rcc}
\hline
~ & {\bf Pump (1557~nm)} & {\bf Stokes (977~nm)}														  \\
\hline
{\bf Mirror Radius 1} & \multicolumn{2}{c}{$\infty$~mm}									\\
{\bf Mirror Radius 2} & \multicolumn{2}{c}{500~mm}											\\
{\bf Zero-Expansion Temperature} & \multicolumn{2}{c}{35~$^{\circ}$C}	  \\
{\bf Length} & 100.13958(7)~mm & 100.15369(7)~mm												\\
{\bf Free Spectral Range} & 1496.873(1)~MHz & 1496.662(1)~MHz						\\
{\bf Finesse}	& $1.37(6)\times10^{4}$ & $1.19(6)\times10^{4}$						\\
{\bf Mode Linewidth} & 109(5)~kHz & 126(5)~kHz													\\
\hline
\end{tabular}
\caption{Key properties of the single ultra-low expansion cavity (ATFilms) to which both the pump and Stokes lasers are referenced. The data presented in this table are extracted from the results presented in figure~\ref{fig:FSR}.}
\label{table:Cavity}
\end{table}

Each beam sent to the cavity passes through an optical fibre-coupled electro-optic modulator (EOM). The output of each EOM (Thorlabs LN65S-FC for 1557~nm, EOSpace PM-0K5-10-PFA-PFA-980 for 977nm) is then coupled via a fibre and mode-matching optics to the optical cavity. Dichroic mirrors at either end of the cavity (Thorlabs BB1-E03P 750 - 1100 nm, and Thorlabs DMLP1180R) are used to combine the two different wavelengths of light entering the cavity, and to separate the two wavelengths following transmission or reflection. The transmitted and reflected beams are monitored on photodiodes, and the signal generated by the reflected light is sent to the locking electronics. The full optical setup is shown in figure~\ref{fig:StirapLayout}~(a) and~(b). The frequency stabilization electronics are a standard Pound-Drever-Hall (PDH) setup as has been explained in \cite{Black:2001}, where the EOM is driven at a frequency~$\omega_{\mathrm{PDH}}\sim10$~MHz to generate the PDH readout signal (figure~\ref{fig:StirapLayout}~(d)). Each laser is fitted with a fast analogue servo module (Toptica FALC~110) to which the produced error signal is sent.

The fibre-coupled EOMs are crucial to the simplicity and flexibility of our setup. These modulators are non-resonant and hence work over a wide bandwidth of~10~GHz. Additionally, these devices can be driven simultaneously at multiple frequencies and require relatively small driving voltages ($\sim4.5$~V). We use these EOMs to provide continuous tunability of the laser frequency sent to the main experiment. By applying a modulation frequency~$\omega_{\mathrm{Offset}}$ to each EOM we add high-frequency sidebands to the original carrier light (figure~\ref{fig:StirapLayout}~(d)). By stabilizing the frequency of a sideband to a cavity mode, we are then able to precisely tune the frequency of the carrier light by simply changing the modulation frequency,~$\omega_{\mathrm{Offset}}$. Due to the high bandwidth of the EOMs,~$\omega_{\mathrm{Offset}}$ may be larger than the free spectral range of the optical cavity~$\omega_{\mathrm{FSR}}$. Hence, the frequency of the carrier light can be tuned continuously to any point between the modes of the cavity. 

The Pound-Drever-Hall technique used to stabilize the frequency of a sideband to a cavity mode requires further modulation of the light at a frequency~$\omega_{PDH}$. We accomplish this using the same non-resonant EOMs already discussed by combining the sideband offset and PDH modulation frequencies on an RF combiner~(Minicircuits ZFSC-2-2-S+) and driving each EOM at two RF frequencies simultaneously, as shown in figure~\ref{fig:StirapLayout}~(c). 

Isolating the optical cavity from vibrations is typically critical in experimental systems such as this for achieving high efficiency STIRAP. Our cavity is placed on a breadboard on top of a sorbathane mat, which is inside a wooden box lined with sound-proofing foam (30~mm thick). The whole assembly is placed on an optical table (without a vibration isolation platform) in the same room as the main experiment itself. We neglect further isolation in part because we find that the part of the apparatus most sensitive to vibrations is not the cavity itself but instead the EOM and the accompanying fibres. 

\begin{figure}
\vspace{0.2cm}
\centering
\includegraphics[width=\textwidth]{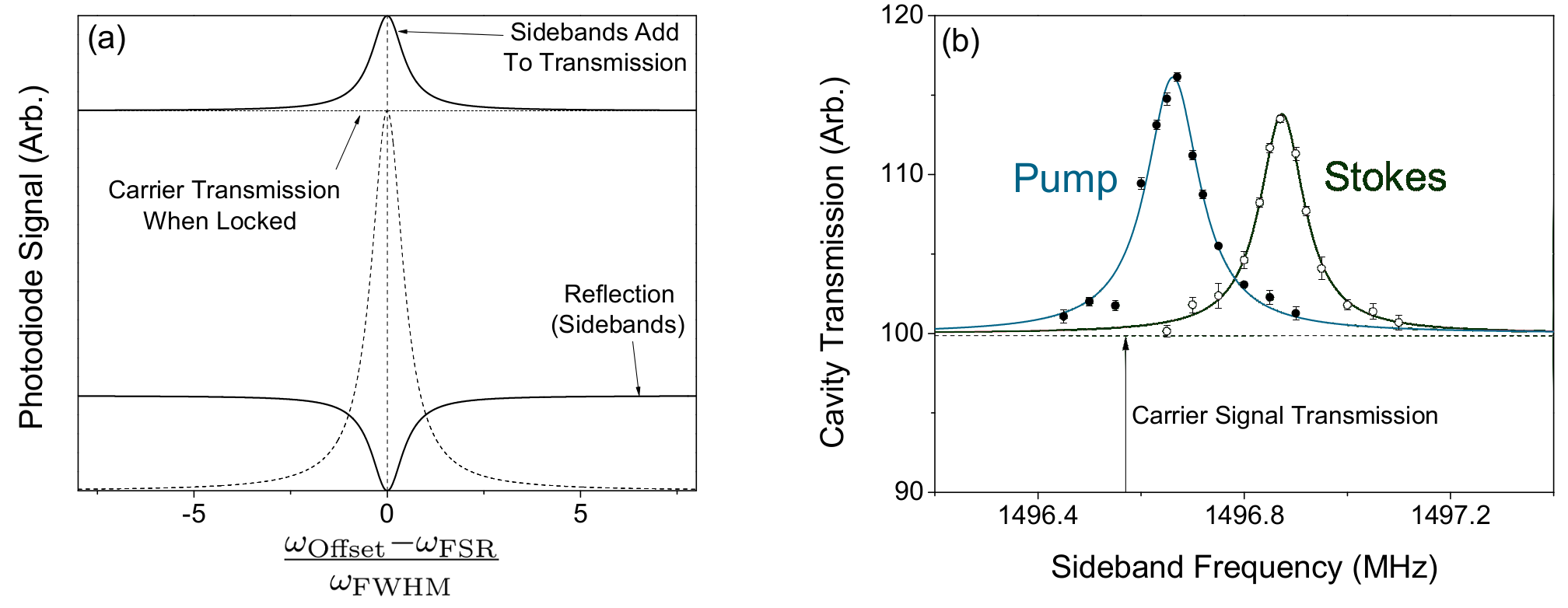}
\caption{Measurement of the free spectral range of the optical cavity, $\omega_{\mathrm{FSR}}$. (a)~The carrier frequency is stabilized to a cavity mode (dashed line). We use a non-resonant EOM to produce sidebands a frequency $\omega_{\mathrm{Offset}}$ away from this cavity mode. Light in the sidebands is reflected by the cavity, except when $\omega_{\mathrm{Offset}}=N\times\omega_{\mathrm{FSR}}$ where $N$ is an integer. In this case, the sideband light is transmitted along with the carrier light and we observe increased transmission through the cavity. By measuring the frequency at which we achieve peak transmission through the cavity as we sweep the sideband across a neighbouring cavity mode (such that $N=1$), we can therefore directly measure $\omega_{\mathrm{FSR}}$. (b)~Experimental measurement of the cavity free spectral range using this method. Results for the cavity transmission are shown with Lorentzian fits at wavelengths of 1557nm (filled circles) and 977nm (empty circles). As the linewidth of our lasers is $\sim$3 orders of magnitude less than the cavity linewidth, we can extract the linewidth of the cavity at each wavelength from the width of the Lorentzian fits.}
\label{fig:FSR}
\end{figure}

Measurement of the absolute wavelengths of each laser is performed using a single wavemeter (Bristol 621A) to which both lasers are coupled. This wavemeter has been calibrated through comparison to the $5P_{3/2} \leftrightarrow 4D_{5/2}$ transition in Rb at 1529~nm, which shows the accuracy of the wavemeter is limited to around~20~MHz. The recent installation of an optical frequency comb as part of a collaboration between Toptica Photonics and Durham University~\cite{Hellerer:2014} should allow for much greater accuracy to be obtained in the near future. Additionally, the accuracy with which we can measure the relative frequency between transitions found using this laser system is only limited by the uncertainty with which we can measure the free spectral range of the cavity. Fortunately, the fibre-coupled EOMs provide a simple, yet accurate method of measuring this quantity. By stabilizing the frequency of the carrier light to a cavity mode, the addition of sidebands which are not resonant with a cavity mode reduces the light transmitted through the cavity. However, if we set the offset modulation frequency such that the sideband overlaps with an adjacent cavity mode, the light in the sideband will once again be transmitted through the cavity. We hence scan the offset modulation frequency and monitor the intensity of the light transmitted through the cavity to measure the position of peak transmission, as shown in figure~\ref{fig:FSR}. As the laser frequency is stabilized to a cavity mode throughout the measurement, the linewidth of the laser is narrowed to $\sim3$ orders of magnitude less than the cavity linewidth (see section~\ref{sec:DSHI}). The width of the transmission peak observed therefore yields the linewidth of the cavity. A Lorentzian fit to the data allows the measurement of both the free spectral range and the linewidth of the cavity at each wavelength, as documented in table~\ref{table:Cavity}. We note that the length of the cavity differs by~14.1(1)~$\mu$m between the two wavelengths, corresponding to the thickness of the inner (977nm) coating on the cavity mirrors.

\begin{figure}
\centering
\includegraphics[width=\textwidth]{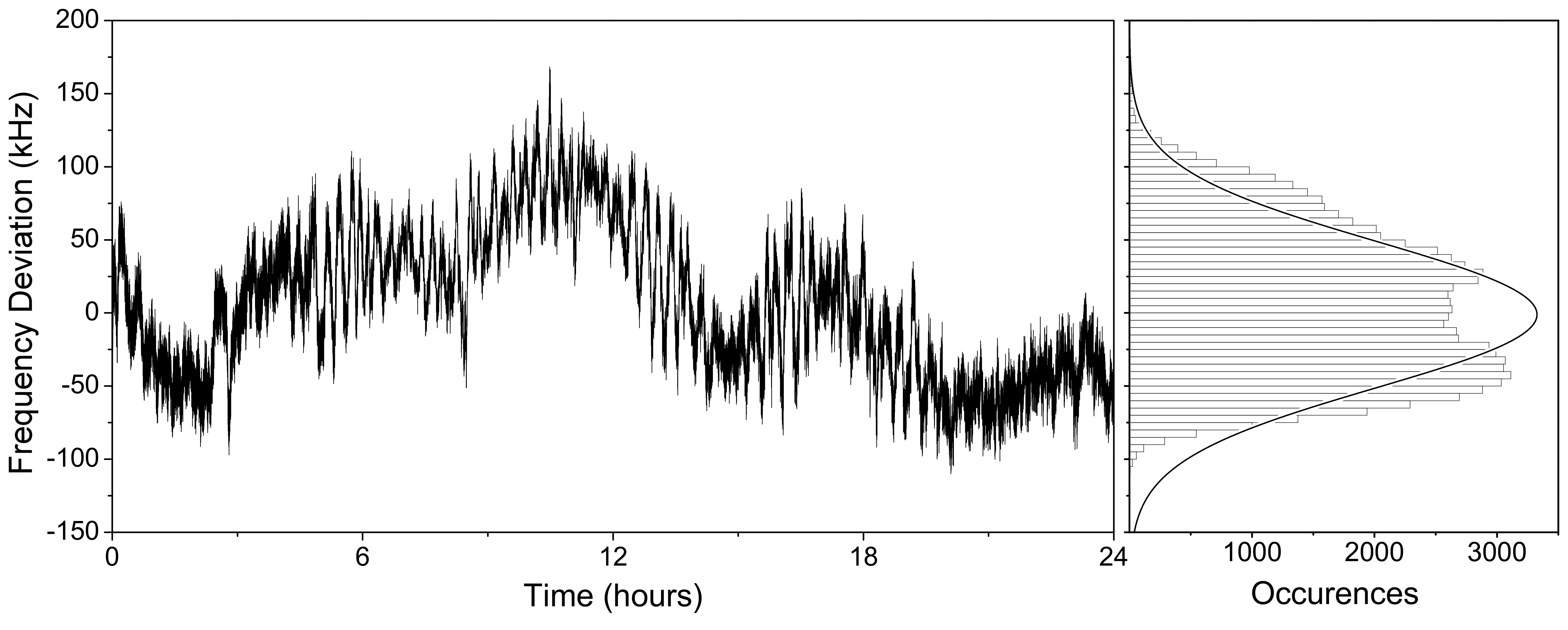}
\caption{Measurement of the stability of the laser by reference to an optical frequency comb. (a) Frequency deviation of the beat note between the pump laser system and a frequency comb tooth, recorded on a counter over a 24-hour period. (b) Histogram of the same data. A Gaussian curve with a full-width half-maximum of 120~kHz is shown for comparison.}
\label{fig:FrequencyComb}
\vspace{0.2cm}
\end{figure}

Long term stability of the laser frequency has been tested by reference to the aforementioned optical frequency comb. This has revealed a root mean square deviation in the frequency deviation of the beat signal over a time period of 24~hours of 116~kHz, as shown in figure~\ref{fig:FrequencyComb}. It is worth noting that in our experiment, it typically takes $1-2$~hours to map out a molecular transition and we have been able to observe transitions with widths of~$\sim200$~kHz.

For molecular spectroscopy and STIRAP, we generate pulses of light at each wavelength by passing each beam destined for the main experiment through separate acousto-optic modulators (AOM). The AOMs (ISOMET 1205C-1023 for 1557~nm, ISOMET 1205C-1 for 977~nm) are driven at their centre frequency of 80~MHz and the first order diffracted light is coupled into 8m single mode polarization maintaining optical fibres providing light to the main experiment. By controlling the amplitude of the AOM driving frequency using a signal supplied by an arbitrary function generator (Agilent~33522B), the power diffracted into the first order may be controlled to create pulses of arbitrary shape. To improve the stability of the AOM response, the AOMs are kept active during the rest of the experimental cycle and the light is instead blocked by a shutter. After the fibres, the two beams are combined on a dichroic mirror (Thorlabs DMLP1180L) and focussed (f = 300~mm) to a waist of 37.7(1)~$\mu$m (pump) and 35.6(6)~$\mu$m (Stokes) at the position of the trapped molecules. The system provides up to 16~mW of each wavelength of light at the position of the molecular sample.

\section{Delayed Self-Heterodyne Measurement of the Laser Linewidth}\label{sec:DSHI}

\begin{figure}
\centering
\includegraphics[width=\textwidth]{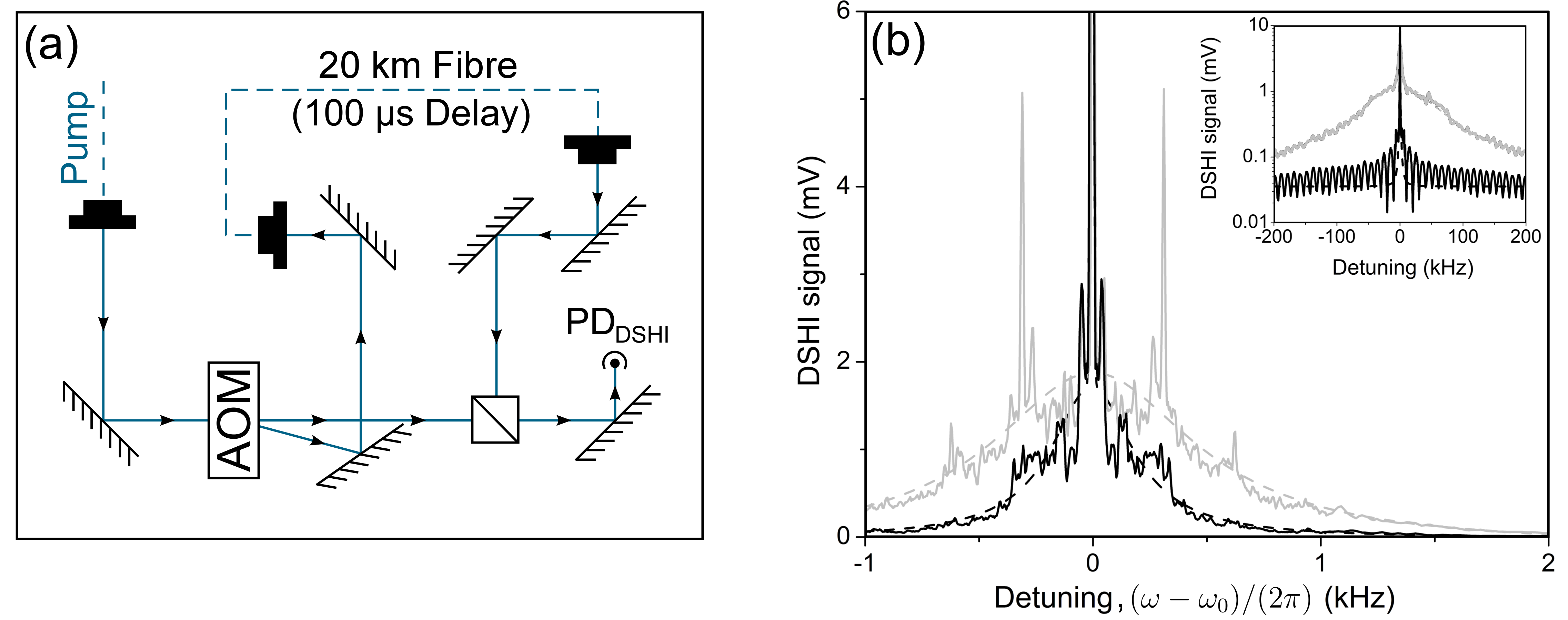}
\caption{Delayed self-heterodyne measurement of the laser linewidth. (a)~Optical layout used for the self-heterodyne linewidth measurement. (b)~Self-heterodyne beatnote for the pump laser. Black and grey lines show linewidths of 0.21(1)~kHz and 0.52(2)~kHz achieved with the main experiment turned off and on respectively. Two additional large spikes at $\pm310$~Hz can be observed when the experiment is switched on indicating the presence of acoustic noise. Inset: the same data over a wider frequency range with a logarithmic amplitude scale, showing the characteristic DSHI interference fringes (black), and the linewidth of the free-running laser (grey) for comparison. This measurement is relatively insensitive to noise below 10~kHz because of the limited delay line length of 100~$\mu$s.}
\label{fig:DSHI}
\end{figure}

We estimate the linewidth of the pump laser using delayed self-heterodyne interferometry~(DSHI)~\cite{Okoshi:1980}. In this method, laser light from the system is separated into two paths, and one of these paths is frequency shifted and time delayed with respect to the other. The paths are then recombined to create a beat note, from which we can estimate the loss of phase coherence. To achieve this, we deliver light from the pump laser through a 2~m single mode fibre to an AOM driven at an angular frequency~$\omega_{0}=2 \pi \times 80$~MHz. The first order diffracted light from the AOM is delayed by $100~\mu$s by a 20~km single mode fibre before being recombined with the light from the zeroth order. The resultant beat note is measured on a high-speed photodiode as shown in figure~\ref{fig:DSHI}~(a). 

In an idealized DSHI experiment, the delay time~$\tau$ between the two paths should be significantly more than the coherence time of the laser~$\tau_{\mathrm{c}}$ such that noise in the two arms of the interferometer are completely uncorrelated~\cite{Richter:1986}. However, it is still possible to get useful information about the linewidth even when $\tau<\tau_{\mathrm{c}}$. In this case, the beat note measured on the photodiode following recombination contains lineshapes resulting from both correlated and uncorrelated noise contributions. If we consider a laser with a constant (white) noise spectrum and power $P_{0}$, this DSHI power spectrum has the analytic form~\cite{Richter:1986}
\begin{eqnarray}\label{eq:DSHIspectrum}
	S_{\mathrm{DSHI}}\left(\omega, \tau\right)&=&\frac{\frac{1}{2}P_0^2\tau_{\mathrm{c}}}{1+(\omega-\omega_{0})^2\tau_{\mathrm{c}}^2}\left(1-e^{-\tau/\tau_{\mathrm{c}}}\left[\cos(\omega-\omega_{0})\tau+\frac{\sin(\omega-\omega_{0})\tau}{(\omega-\omega_{0})\tau_{\mathrm{c}}}\right]\right)\nonumber\\
	&+&\frac{1}{2}P_0^2\pi e^{-\tau/\tau_{\mathrm{c}}}\delta(\omega-\omega_{0}).
\end{eqnarray}
The structure of this spectrum consists of a Lorentzian with full-width half-maximum defined by $1/(2\pi\tau_{c})$ superimposed with interference fringes with a period $1/(2\pi\tau)$ arising from partial coherence between the two paths, and a $\delta$-function at the AOM frequency. For the limiting case of a long delay time where no coherence remains between the paths, $\tau/\tau_{c}\rightarrow\infty$, the power spectrum is simply a Lorentzian curve whose width is set by the laser linewidth. While for $\tau/\tau_{c}\rightarrow0$ this reduces to a $\delta$-function as frequency fluctuations between the two paths become perfectly correlated.

In our system the laser is frequency stabilized to a cavity, so below the servo loop bandwidth the phase noise of the laser is suppressed. This non-uniformity means that we cannot assume that the white noise model gives a good estimate of the coherence time. However, numerical simulations by Di Domenico $et~al.$~\cite{DiDomenico:2010} have shown that the lineshape of such a laser is still approximately Lorentzian, and comes from the part of the individual frequency noise components of each path which exceeds $8\ln(2)/\pi^{2}$ multiplied by their respective Fourier frequencies, known as the $\beta$-separation line. Other parts of the noise spectrum contribute to a wide pedestal without affecting the full-width half-maximum of the lineshape. Hence, a reasonable estimate of the linewidth can still be achieved using the same functional form as equation~\ref{eq:DSHIspectrum}.

The measured self-heterodyne beat note of the laser (with the AOM frequency removed) is shown in figure \ref{fig:DSHI}. The inset shows the same signal over a larger span with interference fringes as predicted in equation \ref{eq:DSHIspectrum}. For our setup we expect $\tau/\tau_{c}\sim 0.1$ and around this value the oscillatory term is reasonably flat in the range of 2~kHz from the centre. We also note that in our measurement, the $\delta$-function will be broadened due to the limited resolution bandwidth ($R$) of our spectrum analyser (Agilent N9320B). The $\delta$-function can therefore be replaced with an appropriately normalized Gaussian and the fringes can be neglected:
\begin{eqnarray}\label{eq:DSHIspectrumred}
	S_{\mathrm{DSHI}}\left(\omega\right)&=&\frac{\frac{1}{2}P_0^2\tau_c}{1+(\omega-\omega_{0})^2\tau_c^2} + \sqrt{\frac{\pi}{8}} \frac{P_{0}^{2}}{R} \exp\left[- \frac{1}{2} \left( \frac{\omega-\omega_{0}}{R}\right) ^{2}\right].
\end{eqnarray}
Fitting to this equation over a range of 2~kHz suggests a laser linewidth of 0.21(1)~kHz. However, when the equipment used in the rest of the experiment was turned on the linewidth increases to 0.52(2)~kHz. The main contribution to this noise is acoustic, coming from the large power supplies used to drive the magnetic field coils for the experiment, with smaller contributions from the water cooling pump and the fibre laser used in the optical dipole trap. For comparison, the free-running laser linewidth is measured to be 85(8)~kHz as shown inset in figure~\ref{fig:DSHI}~(b). This measurement is much closer to the idealized DSHI case as the coherence time of the laser is much shorter. 

An independent analysis of the DSHI spectrum for the same experimental setup was performed using proprietary phase noise reconstruction software~\cite{Puppe:2014}. This yielded a linewidth around 200~Hz which agrees well with the reading from the spectrum analyser. It should be noted neither analysis is a full measurement of the linewidth since the 100~$\mu$s delay in the DSHI method acts as a 10~kHz high-pass filter, reducing sensitivity at low frequencies. A full measurement by comparison with a second identical system would be prohibitively expensive.

\section{Producing Ultracold Atomic Mixtures}\label{sec:Experimental}

\begin{figure}
\centering
\includegraphics[width=\textwidth, trim=0.5cm 1cm 0.75cm 0.5cm]{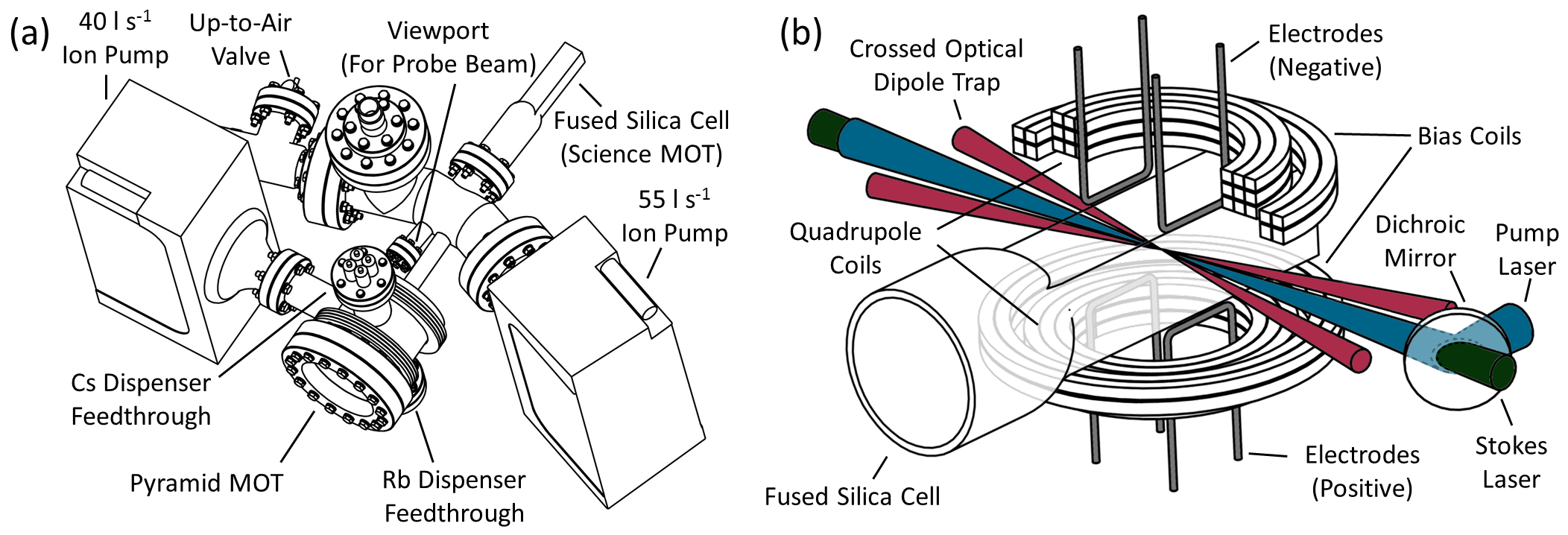}
\caption{(Colour Online) Experimental apparatus for the creation of $^{87}$Rb$^{133}$Cs molecules. (a) The vacuum system consisting of two MOTs. A pyramid MOT acts as a cold dual-species atom source for the second labelled as the science MOT. (b) The science MOT, with the locations of field coils, STIRAP lasers and electrodes shown.}
\label{fig:Apparatus}
\end{figure}

Details of our experiment have been described extensively in the context of our studies of dual-species condensates~\cite{McCarron:2011, Cho:2011}. The experimental apparatus consists of two magneto-optical traps (MOTs). The first, a pyramid MOT, acts as a dual-species cold atom source for the second, referred to as the science MOT~\cite{Harris:2008}. The vacuum system layout and science MOT apparatus are shown in figure~\ref{fig:Apparatus}~(a) and~(b), respectively. Following trapping in the science MOT, the $^{87}$Rb and $^{133}$Cs atoms are optically pumped to the $\ket{F=1, m_{F}=-1}$ and $\ket{3, -3}$ low-field-seeking states respectively and subsequently transferred into a magnetic quadrupole trap. The $^{87}$Rb atoms are further cooled by forced RF evaporation while interspecies elastic collisions cool the $^{133}$Cs atoms sympathetically until Majorana losses~\cite{Lin:2009} limit further evaporation. We then load the atoms into a magnetically levitated crossed optical dipole trap~\cite{Jenkin:2011}, using RF adiabatic rapid passage to transfer the $^{87}$Rb and $^{133}$Cs atoms into the $\ket{1, 1}$ and $\ket{3,3}$ high-field-seeking states respectively. The combination of bias field and quadrupole field is chosen to levitate both of these states against gravity. By reducing the optical trap powers, we routinely evaporate to a nearly-degenerate sample of $\sim 2.5 \times 10^{5}$~$^{87}$Rb atoms and $\sim 2.0 \times 10^{5}$~$^{133}$Cs atoms at a temperature of $\sim 300$~nK, from which we can begin magnetoassociation. 

We have recently added a means of applying electric fields in our apparatus using the array of four electrodes shown in figure~\ref{fig:Apparatus}~(b). Each steel electrode is 1.5~mm in diameter and 22.0(5)~mm long, and they are separated by 29.0(2)~mm vertically and 24.8(2)~mm horizontally. Applying an electric potential of 1~kV between the upper and lower electrode pairs yields an electric field at the position of the molecules of 153(1)~V~cm$^{-1}$~\cite{Molony:2014}. We are able to apply 6.5~kV before polarization of the fused silica cell becomes detectable, limiting the maximum electric field accessible in the experiment to~1~kV~cm$^{-1}$.

\section{Creating Ultracold $^{87}$Rb$^{133}$Cs Molecules}\label{sec:Molecules}

\begin{figure}
\centering
\includegraphics[width=\textwidth]{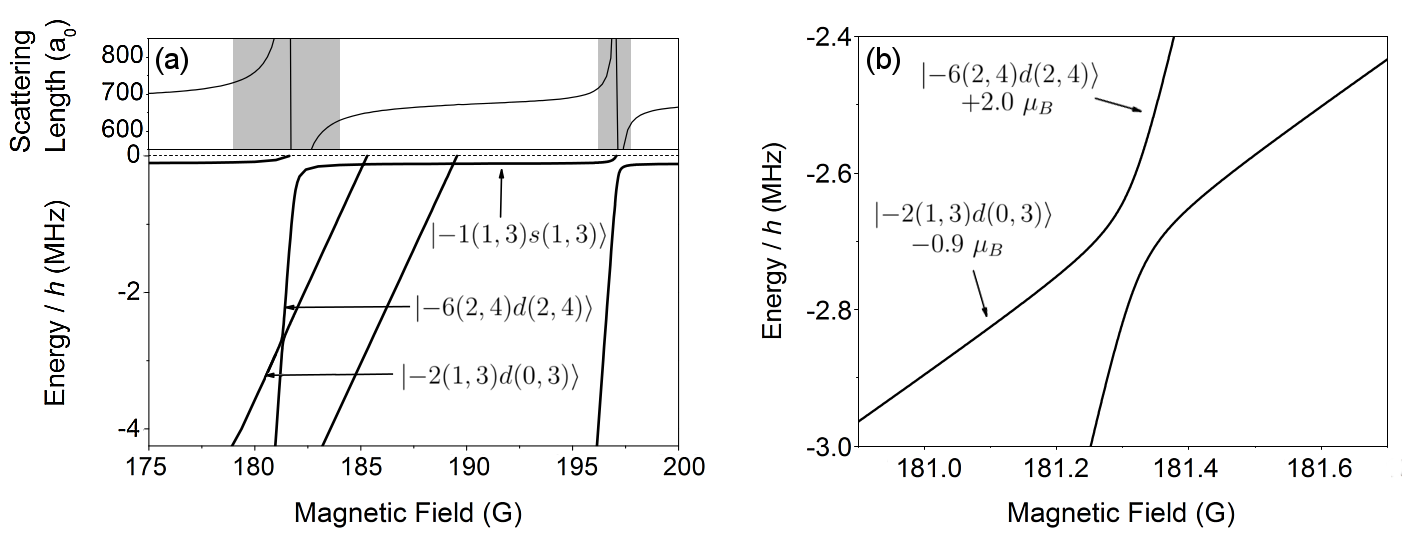}
\caption{Near-threshold $^{87}$Rb$^{133}$Cs molecular states. (a) Upper panel: the interspecies scattering length between $^{87}$Rb and $^{133}$Cs, the grey regions highlight the locations of the two Feshbach resonances in the relevant magnetic field region. Molecules are associated on the higher-field resonance at 197~G. Lower panel: Weakly-bound states relevant to the Feshbach association sequence. (b) The avoided crossing between the $\ket{-6(2,4)d(2,4)}$ and $\ket{-2(1,3)d(0,3)}$ states which is followed both during the association sequence and following the transfer to a pure optical dipole trap for ground state transfer.}
\label{fig:Feshbach}
\end{figure}

The near-threshold bound states relevant for the magnetoassociation of $^{87}$Rb$^{133}$Cs molecules are shown in figure~\ref{fig:Feshbach}(a). These states are labelled as $\ket{n(f_{\mathrm{Rb}},f_{\mathrm{Cs}})L(m_{f_{\mathrm{Rb}}},m_{f_{\mathrm{Cs}}})}$, where $n$ is the vibrational label for the particular hyperfine $(f_{\mathrm{Rb}},f_{\mathrm{Cs}})$ manifold, counting down from the least-bound state which has ${n=-1}$, and $L$ is the quantum number for rotation of the two atoms about their centre of mass, following the convention laid out in~\cite{Takekoshi:2012}. Note that all states have $M_{\mathrm{tot}}=4$, where $M_{\mathrm{tot}} = M_{F} + M_{L}$ and $M_{F} = m_{f_{\mathrm{Rb}}} + m_{f_{\mathrm{Cs}}}$. Magnetoassociation is performed in the magnetically levitated crossed dipole trap by sweeping the bias field down across a Feshbach resonance at 197.10(3)~G with a speed of 250~G~s$^{-1}$ to produce molecules in the $\ket{-1(1,3)s(1,3)}$ state. The bias field is then reduced rapidly, to transfer the molecules into the $\ket{-2(1,3)d(0,3)}$ state at 180.487(4)~G via the path shown in figure~\ref{fig:Feshbach}(a). The magnetic quadrupole field required to levitate the molecules in this state causes the remaining atoms to be over-levitated, which allows for purification of the molecular cloud via the Stern-Gerlach effect. The number of molecules we produce is optimized by varying the ratio of $^{87}$Rb and $^{133}$Cs before association by changing the number of Cs atoms loaded into the science MOT. We find that the molecule production is maximized when the mean phase-space density of the mixture is maximized (see figure~5 in~\cite{Koppinger:2014}). To detect the molecules, we quickly ramp back across the same resonance to dissociate into atoms. Both species of atoms are detected by absorption imaging, with the probe light propagating along the axis of the cell. We typically create trapped samples of $\sim 2500$~molecules with the same temperature as the original atomic sample, and a lifetime in the $\ket{-2(1,3)d(0,3)}$ state of 200~ms. We attribute the low conversion efficiency to the large interspecies scattering length of $\sim 650~a_{0}$~\cite{Takekoshi:2012, McCarron:2011}, which limits the phase-space densities of the atomic samples.

Populating the $\ket{-2(1,3)d(0,3)}$ state is convenient in our apparatus, as the quadrupole field which we generate can only magnetically levitate states with a negative magnetic dipole moment. However, population of the neighbouring weak-field seeking~$\ket{-6(2,4)d(2,4)}$ state is still possible by transferring the molecules into a pure optical trap which is sufficiently deep to support the molecules against gravity. This is achieved through a three stage process. First, the dipole trap power is increased from 200~mW to 1~W, increasing the trap depth to 12.7~$\mu$K. The magnetic levitation gradient is then removed, followed by a bias field ramp up to 181.624(1)~G at a speed of 2.3~G~ms$^{-1}$. The critical ramp speed ($\dot{r_{c}}$), below which the avoided crossing between two states (shown in figure~\ref{fig:Feshbach}(b)) is adiabatically followed is given by,
\begin{equation}
\dot{r_{c}} = \frac{2 \pi V^{2}}{\hbar {\Delta \mu}},
\end{equation}
where $V$ is the coupling strength, and $\Delta \mu$ is the difference in the magnetic moment between the two states. The critical ramp speed for this avoided crossing is $\sim70$~G~ms$^{-1}$, significantly greater than the ramp speed we use. Hence, we observe the molecules being transferred efficiently between the two states. The transfer to a tighter dipole trap heats the molecules to a temperature of 1.5~$\mu$K, due largely to the adiabatic compression of the gas. However, it has the advantage of removing the variable Zeeman shift across the cloud which results from the use of a magnetic field gradient. We observe a reduction in the lifetime of the molecules from 0.21(1)~s in the~$\ket{-2(1,3)d(0,3)}$ state to 23(2)~ms in the~$\ket{-6(2,4)d(2,4)}$.

\section{Molecular Spectroscopy}\label{sec:MolecularSpectroscopy}

To implement STIRAP, the pump and Stokes lasers need to be tuned such that they couple the initial weakly-bound state and the rovibrational ground state to a common excited state. The transitions to this common state therefore need to be in a range accessible to both lasers. In our system this corresponds to a range of states lying between $\sim6390$~cm$^{-1}$ and $\sim6540$~cm$^{-1}$ above the dissociation energy of the molecule as shown in figure~\ref{fig:MolecularPotentials}. Detailed spectroscopy of the mixed $A^{1}\Sigma^{+}+b^{3}\Pi$ molecular potential by Debatin~$et~al.$~\cite{Debatin:2011} has already identified the lowest hyperfine sublevel of the~$\ket{E}\equiv\ket{\Omega'=1, v'=29, J'=1}$ state as suitable for efficient ground state transfer in this region. To demonstrate the capabilities of our laser system, we perform molecular loss spectroscopy on seven of the electronically excited states previously identified including this state.
 
\begin{figure}
\centering
\includegraphics[width=\textwidth]{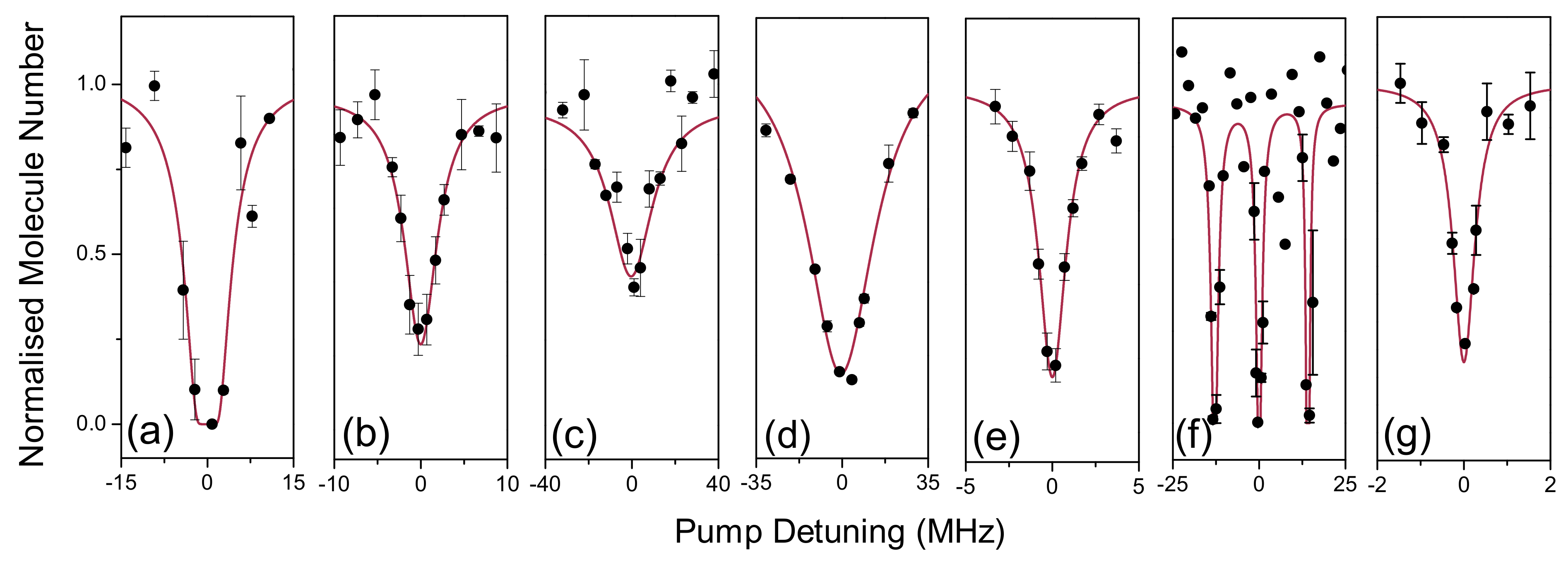}
\caption{One photon molecular spectroscopy from the $\ket{-2(1,3)d(0,3)}$ state close to dissociation. Experimental results showing the observed transitions to molecular states in the $A^{1}\Sigma^{+}+b^{3}\Pi$ hyperfine manifold. The pump detuning in each case is relative to the centre of the transition. The states are as labelled as in table~\ref{table:Spectroscopy} where the absolute wavelength measured for each transition is presented.}
\label{fig:OnePhoton}
\end{figure}

Molecular loss spectroscopy is carried out by illuminating the Feshbach molecules with a 750~$\mu$s pulse of pump light, polarized parallel to the magnetic field. Pump light resonant with a transition to an excited molecular state leads to a reduction in the number of molecules in the trap. Starting with the maximum power of $\sim16$~mW, the power in the pump beam is reduced until a small number of molecules is still observable even when directly on resonance in order to get an accurate measure of the transition centre. A number of these loss features may be seen in figure~\ref{fig:OnePhoton}; each feature is recorded using a different power in the pump beam, ranging between 300~$\mu$W to 16~mW, due to the variation in coupling strengths between states. Each state is found with a constant bias field of~180.487(4)~G applied to the molecules, to initialize them in the~$\ket{-2(1,3)d(0,3)}$ state. To compare the wavelength of each transition to that previously reported~\cite{Debatin:2011}, we must subtract the Zeeman shift of~327~MHz caused by the presence of the bias field from the measured wavelength for each transition. These values are presented in table~\ref{table:Spectroscopy}. 

\begin{table}
\centering
\begin{tabular}{c c c c}
\hline 
 & {\bf State} & \multicolumn{2}{c}{\bf Transition Energy $E/hc$ (cm$^{-1}$)} \\
 & 						 & {\bf Innsbruck~\cite{Debatin:2011}}	& {\bf Durham} \\
\hline
(a) & $\ket{\Omega' = 0, v' = 35, J' = 1}$ & 6364.031(2) & 6364.0301(7) \\
(b) & $\ket{\Omega' = 0, v' = 37, J' = 1}$ & 6398.663(2) & 6398.6584(7) \\
(c) & $\ket{\Omega' = 0, v' = 38, J' = 1}$ & 6422.986(2) & 6422.9730(7) \\
(d) & $\ket{\Omega' = 0, v' = 38, J' = 3}$ & $-$ & 6423.1149(7) \\
(e) & $\ket{\Omega' = 1, v' = 29, J' = 1}$ & 6423.501(2) & 6423.5026(7) \\
(f) & $\ket{\Omega' = 1, v' = 29, J' = 2}$ & $-$ & 6423.5843(7) \\
(g) & $\ket{\Omega' = 1, v' = 29, J' = 3}$ & $-$ & 6423.6847(7) \\
\hline
\end{tabular}
\caption{Table detailing all seven excited states studied, and the transition energy ($E/hc$) of each. Note that although we can excite the $\ket{\Omega' = 0, v' = 35, J' = 1}$ state transition with the pump laser, it lies outside of the region accessible to the Stokes laser and so it is not possible to use this state as our intermediate STIRAP state.}
\label{table:Spectroscopy}
\end{table}

As discussed in section~\ref{sec:Molecules}, the state we begin our ground state transfer is not limited to the~$\ket{-2(1,3)d(0,3)}$ state. In fact, the~$\ket{\Omega'=1, v'=29, J'=1}$ state we use has much stronger coupling to the~$\ket{-6(2,4)d(2,4)}$ state. We measure the strength of the coupling by varying the duration of the pump pulse used~($t$), and measuring the fraction of the molecules remaining in each of the possible initial states~($N/N_{0}$). The results are then fitted to
\begin{equation}
\frac{N}{N_{0}} = \exp{\left(\frac{-\Omega_{\mathrm{p}}^{2}t}{\gamma}\right)},
\end{equation}
where $\Omega_{\mathrm{p}}$ is the Rabi frequency and $\gamma$ is the natural linewidth of the pump transition. The transition Rabi frequencies are measured as $\Omega_{\mathrm{p}}=2\pi\times17(5)$~kHz from the~$\ket{-2(1,3)d(0,3)}$ state and $\Omega_{\mathrm{p}}=2\pi\times0.18(1)$~MHz from the~$\ket{-6(2,4)d(2,4)}$ state. As the Rabi frequency for the transition from the~$\ket{-6(2,4)d(2,4)}$ state is $\sim10$ times greater than that achieved for the~$\ket{-2(1,3)d(0,3)}$ state transition, we hence use~$\ket{F}\equiv\ket{-6(2,4)d(2,4)}$ for two-photon experiments.

\begin{figure}
\centering
\includegraphics[width=\textwidth]{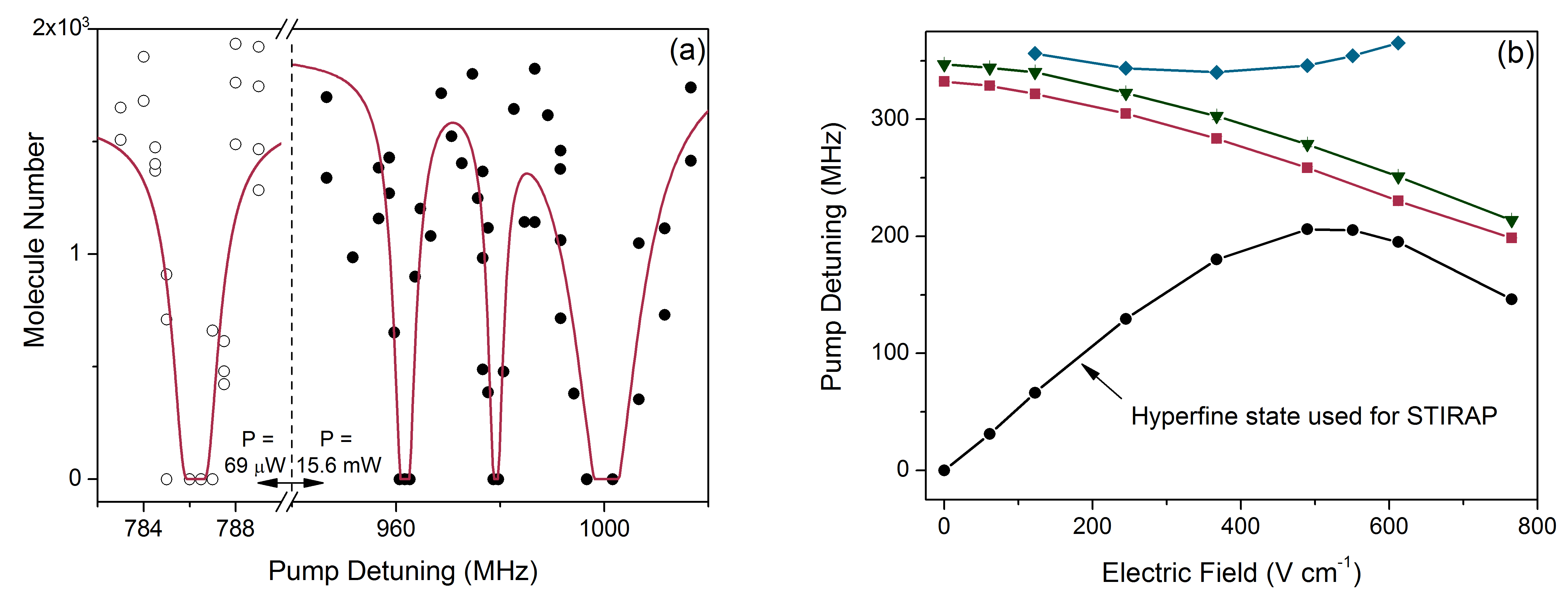}
\caption{One photon Stark spectroscopy from the $\ket{-6(2,4)d(2,4)}$ state close to dissociation. (a)~Spectroscopy of the four hyperfine sublevels of the~$\ket{\Omega' = 1, v'=29, J'=1}$ state observed with an applied electric field of 245~V~cm$^{-1}$. Empty circles represent data collected with a pump power of~69~$\mu$W, filled circles signify a pump power of~$\sim15.6$~mW. (b)~Stark shift of these states up to an applied electric field of~765~V~cm$^{-1}$. The lowest hyperfine sublevel is used for STIRAP transfer to the rovibrational ground state. An avoided crossing is observed between that state and the higher-lying hyperfine states at an applied electric field of $\sim 550$~V~cm$^{-1}$.}
\label{fig:ExcitedStarkSpectroscopy}
\end{figure}

We further characterize the~$\ket{\Omega'=1, v'=29, J'=1}$ state by measuring its DC Stark shift. By applying a 750~$\mu$s pulse of pump light whilst an electric field is applied to the molecules we can track the DC Stark shift of the various hyperfine sublevels of the state from their zero field values as shown in figure~\ref{fig:ExcitedStarkSpectroscopy}. The DC Stark shift of the state is initially linear with a gradient of 500~kHz/(V cm$^{-1}$) up to $\sim400$~V cm$^{-1}$. Above this field, we observe an avoided crossing between the sublevel identified for STIRAP and the higher-lying hyperfine states. It is worthy of note that the coupling to these higher-lying states is relatively weak as $\sim15.6$~mW of available pump power is required to saturate the transitions, whereas only 69~$\mu$W is necessary to saturate the lowest hyperfine sublevel which we use for STIRAP (see figure~\ref{fig:ExcitedStarkSpectroscopy}~(a)). Measurement of this Stark shift was crucial in obtaining our recent measurement of the molecule-frame dipole moment of ground state $^{87}$Rb$^{133}$Cs as~1.225(3)(8)~D, where the values in parentheses are the statistical and systematic uncertainties respectively~\cite{Molony:2014}.

\begin{figure}
\centering
\includegraphics[width=0.5\textwidth]{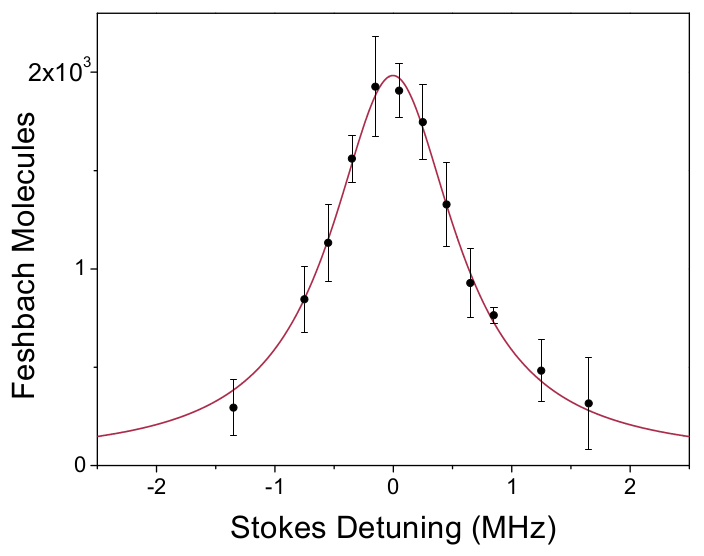}
\caption{Two-photon spectroscopy of the rovibrational ground state. The pump laser is resonant with the transition between the $\ket{F}$ and $\ket{E}=\ket{\Omega'=1, v'=29, J'=1}$ states, and is left on continuously while the Stokes laser is pulsed on. The molecules remain in the initial near-dissociation state when the Stokes light is on resonance with the ground-state transition due to the formation of a dark state~\cite{Lang:2008b}. The fitted curve is Lorentzian with a full-width half-maximum of 1.38(9)~MHz. This state was determined to be the $\ket{v''=0, J''=0}$ state via measurement of the rotational splitting between this state and the $\ket{v''=0, J''=2}$ state~\cite{Molony:2014}. }
\label{fig:TwoPhoton}
\end{figure}

To detect low lying molecular levels of the singlet potential we use two-photon dark-state spectroscopy~\cite{Lang:2008b, Mark:2009}. This is performed by keeping the pump laser on resonance with the $\ket{F}$ to $\ket{E}$ transition, and pulsing both the pump and Stokes lasers on simultaneously for 750~$\mu$s. The Stokes power is set to the maximum available of 16~mW, while the power of the pump laser is set to 40~$\mu$W such that the pump transition is slightly saturated. When the Stokes light is off resonance, we therefore observe no molecules after the pulse. If the Stokes light is on resonance with a transition to state~$\ket{E}$ however, the molecules are projected onto the dark state given by equation~\ref{eqn:DarkState}. Imaging the dissociated atoms after the pulse sequence corresponds to a projection of this dark state back onto the initial Feshbach state. This double projection results in a final state given by $\cos^{2}(\theta)\ket{F}$. During the pulse, the Stokes Rabi frequency is much higher than the pump Rabi frequency and hence the mixing angle $\theta$ is small. This leads to a large proportion of the Feshbach state remaining following the pulse sequence, which we observe as a suppression of the molecular loss. This method allowed the search for the rovibrational ground state before attempting STIRAP. The result of scanning the Stokes frequency across the transition to the ground state is shown in figure~\ref{fig:TwoPhoton}.


\section{STIRAP Transfer to the Rovibrational Ground State}\label{sec:GroundStateTransfer}

\begin{figure}
\centering
\includegraphics[width=\textwidth]{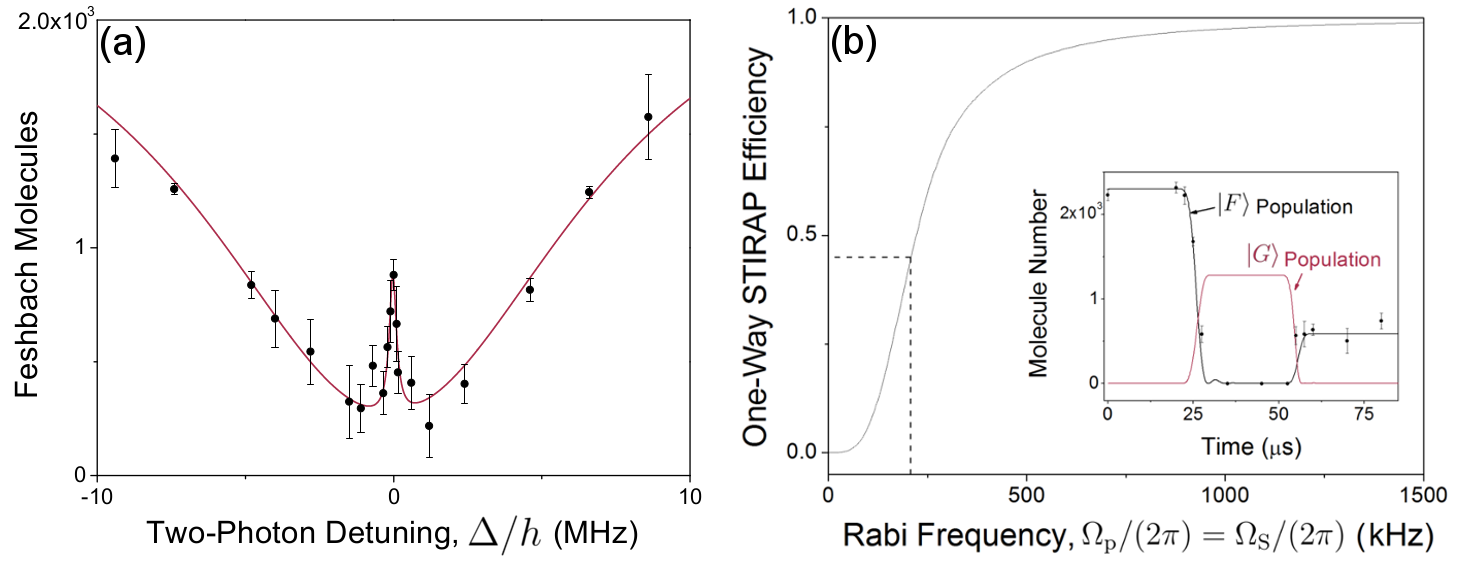}
\caption{(Colour Online) Transfer of molecules close to dissociation to the rovibrational ground state via STIRAP. (a)~A two-way STIRAP pulse is used as shown in figure~\ref{fig:MolecularPotentials}~(c). Both lasers are detuned by the same frequency $\Delta/h$ away from the resonant frequency for their respective transitions. At large detunings molecules never leave the initial state. (b) Main: The STIRAP efficiency dependence upon Rabi frequency. The Rabi frequency of the pump and Stokes transitions are set to be equal. The dashed line shows the expected efficiency at the observed maximum pump Rabi frequency of $2\pi\times0.18$~MHz. Inset: The experimentally measured population in the initial state~$\ket{F}$ at various times during the pulse sequence. The solid black line is generated by numerical simulations (see text). The pump Rabi frequency is set to the experimentally measured value, while the Stokes Rabi frequency is fitted as a free parameter. The red dashed line illustrates the population of the ground state~$\ket{G}$.}
\label{fig:STIRAP}
\end{figure}

To transfer the molecules to the rovibrational ground state, we use a pulse sequence which begins with the Stokes beam at a power of 7~mW initially turned on for 20~$\mu$s. The power of each laser is then ramped sinusoidally over the next~10~$\mu$s such that the pump laser power becomes $\sim16$~mW and the Stokes laser is reduced to zero, in order to adiabatically transfer the population to the ground state. As we cannot directly image molecules in the ground state, after a~20~$\mu$s hold the ramp sequence is reversed to transfer back to the initial state for dissociation and imaging~(as shown in figure~\ref{fig:MolecularPotentials}~(c)). Figure~\ref{fig:STIRAP}~(a) shows the result of varying the detuning of both the pump and Stokes lasers simultaneously by the same amount across the excited state~$\ket{E}$. At large detunings, both lasers are sufficiently off-resonant with the excited state that the molecules are no longer removed from the initial state~$\ket{F}$. The maximum STIRAP efficiency is achieved when both lasers are exactly on resonance, where a one-way efficiency of $\sim50~\%$ is achieved. This equates to a population of over 1000 molecules in the rovibrational ground state. The time evolution of the initial state population during the pulse sequence has already been discussed in our initial report of ground state transfer~\cite{Molony:2014}, but is shown for completeness here inset in figure~\ref{fig:STIRAP}~(b).

To understand the limiting factors in our transfer efficiency we model the transfer by numerically integrating the Lindblad master equation for the four level scheme shown in figure~\ref{fig:MolecularPotentials}~(b)~\cite{Molony:2014}. Non-adiabaticity of the transfer causes population of the excited state which leads to molecule loss. We model this loss as a decay of state~$\ket{E}$ to a dump level~$\ket{X}$ at a rate determined by the natural linewidth of the excited state~$\gamma$. By setting the Rabi frequency of the pump and Stokes transitions equal, we can see how the magnitude of this common Rabi frequency affects the efficiency of the ground state transfer as shown in figure~\ref{fig:STIRAP}~(b). The Rabi frequency which we measure for the pump transition (see section~\ref{sec:MolecularSpectroscopy}) is much lower than that used in previous work~\cite{Debatin:2011}, and correlates to a low one-way efficiency similar to that which we observe in our experiment. We believe that this is likely due to an offset between the focal point of the STIRAP beam and the position of the molecular cloud, resulting in the molecules experiencing a lower light intensity and hence the transitions are driven at a lower Rabi frequency. Fortunately the shape of the efficiency curve indicates that we are in a region with a large gradient; only a small improvement in the Rabi frequency of each transition should therefore be necessary to yield a large improvement in transfer efficiency. We hope to achieve this in the near future through careful realignment of the STIRAP beam focusing lens.    

\section{Conclusion}\label{sec:Conclusion}
In summary, we have described and characterized a simple, versatile dual-wavelength laser system which can be used to transfer weakly bound Feshbach molecules to the rovibrational ground state. At the heart of the setup is a single ULE cavity to which both lasers are frequency stabilized. The use of non-resonant EOMs allows continuous tunability of the laser frequency between cavity modes, whilst also providing the modulation needed to generate the Pound-Drever-Hall error signal. Additionally, the EOMs provide a simple method for accurately determining the free spectral range of the cavity. The frequency stabilized laser linewidth has been
estimated to be 0.21(1)~kHz by the method of delayed self-heterodyne interferometry. The long-term frequency stability was tested by beating one of the lasers with the output of an optical frequency comb. The rms deviation of the beat signal over a 24~hour period was 116~kHz. We have demonstrated performance of the laser system by performing one and two photon molecular spectroscopy using a sample of ultracold $^{87}$Rb$^{133}$Cs Feshbach molecules. The laser system was then
used to transfer the weakly-bound molecules to the rovibrational ground state with a
one-way efficiency of $\sim$50~\%, generating a sample of over 1000~ground state molecules. We expect that with further careful alignment transfer efficiencies $>90$~\% should be possible. We believe the simple setup outlined in this paper will be useful to the many groups now pursuing ultracold ground state molecules.

\section*{Acknowledgments}

We would like to thank J. M. Hutson, C. R. Le Sueur and C. L. Blackley for many valuable discussions and for providing theoretical-bound state energies and magnetic moments of $^{87}$Rb$^{133}$Cs, Toptica Photonics and T. Puppe for the loan of a 20~km optical fibre and helping in the analysis of the self-heterodyne data, T.~Ogden for help in development of the STIRAP simulation and D. L.~Jenkin, D. J.~McCarron and H. W.~Cho for their work on the early stages of the project. This work was supported by the UK EPSRC [Grants EP/H003363/1, EP/I012044/1 and ER/S78339/01]. The data presented in this paper are available upon request.


\section*{References}

\bibliography{NJP_SpecialIssue_References}

\end{document}